\documentclass[12 pt]{article}

\usepackage{amsmath,amssymb,amsfonts,latexsym}
\usepackage{rotating}

\begin{document}

\newtheorem{dfn}{Definition}[section]
\newtheorem{theorem}{Theorem}[section]
\newtheorem{axiom2}{Example}[section]
\newtheorem{axiom3}{Lemma}[section]
\newtheorem{lem}{Lemma}[section]
\newtheorem{prop}{Proposition}[section]
\newtheorem{cor}{Corollary}[section]
\newcommand{\be}{\begin{equation}}
\newcommand{\ee}{\end{equation}}
\newcommand{\lmat}{\left(\begin{array}{cccccc}}
\newcommand{\rmat}{\end{array}\right)}
\newcommand{\lm}{\lambda}
\newcommand{\al}{\alpha}
\newcommand{\ID}{{\mathbb{D}}}
\newcommand{\IG}{{\mathbb{G}}}
\newcommand{\X}{{\mathbb{X}}}
\newcommand{\Y}{{\mathbb{Y}}}
\newcommand{\p}{\partial}
\newcommand{\bel}{\begin{equation}\label}

\title{On Quantized Li\'{e}nard Oscillator and Momentum Dependent Mass \\
 }

\author {B. Bagchi\footnote{E-mail: {\tt bbagchi123@rediffmail.com}}\\
Department of Applied Mathematics \\
University of Calcutta\\ 92 Acharya Prafulla Chandra Road \\ Kolkata - 700009,  India \\
\and
A. Ghose Choudhury \footnote{E-mail: {\tt aghosechoudhury@gmail.com}}\\
Department of Physics, Surendranath  College,\\ 24/2 Mahatma
Gandhi Road, Kolkata-700009, India.\\
\and
Partha Guha\footnote{E-mail: {\tt partha@bose.res.in}}\\
Institut des Hautes \'Etudes Scientifiques\\
Le Bois-Marie 35, Route de Chartres \\
F-91440 Bures-sur-Yvette, France \\
\and
S.N. Bose National Centre for Basic Sciences \\
JD Block, Sector III, Salt Lake \\ Kolkata - 700098,  India\\}

\date{}

\maketitle

\smallskip

\smallskip

\begin{abstract}
We examine the analytical structure of the nonlinear Li\'enard
oscillator and show that it is a bi-Hamiltonain system depending
upon the choice of the coupling parameters. While one has been
recently studied in the context of a quantized momentum-dependent
mass system, the other Hamiltonian also reflects a similar feature
in the mass function and also depicts an isotonic character. We
solve for such a Hamitonian and give the complete solution in
terms of a confluent hypergeometric function.
\end{abstract}

\paragraph{Mathematical Classification} 70H03, 81Q05.

\smallskip

\paragraph{Keywords and Key phrases} Li\'{e}nard equation, Jacobi last multiplier,  momentum dependent mass function.

\section{Introduction}
Exploring the Schr\"{o}dinger equation in the momentum space is
often advantageous because many quantities of physical interest
are more readily evaluated in this representation rather than in
the coordinate formulation. This is especially true for some
typical scattering problems and form factors of certain kinds
\cite{Taylor, BJ}. It is worthwhile to recall that for the simple
one-dimensional hydrogen atomic system it took well over thirty
years to fully appreciate its underlying principles and that too
after an analysis was carried out in the momentum space
representation \cite{Loud, NYabc}. Very recently an interesting
aspect has been brought to light that concerns the relevance of a
momentum-dependent mass for a quantized nonlinear oscillator of
Li\'{e}nard type \cite{RSL,MRS, GCGK, NT,NL2} \be\label{e1}
\ddot{x}+kx\dot{x}+\omega^2 x+\frac{k^2}{9}x^3=0,\ee where $k$ and
$\omega$ are real parameters.  It admits  a periodic solution for
$x(t)$ namely \be\label{ex3}x=\frac{A\sin(\omega
t+\delta)}{1-\frac{kA}{3\omega}\cos(\omega t+\delta)},\ee where
$A$ and $\delta$ are arbitrary constants subject to
$-1<kA/3\omega<1$. An intriguing feature about the Hamiltonian  of
(\ref{e1}) is that it depicts an interchange of the roles of the
position and momentum variables in its kinetic and potential
energy terms respectively. However, it is important to realize
that its quantization is hard to tackle in the coordinate
representation of the Schrodinger equation but can be successfully
carried out in the momentum space.
Note  that the Hamiltonian tied-up with (\ref{e1}) is not unique:
rather it points to a bi-Hamiltonian system. This is not
surprising since,  it also reveals,
in the quantum case, a Goldman and Krivchenkov-type isotonic
system \cite{Yes,WJ} given by the half-line combination of a harmonic
oscillator and centrifugal barrier-like term. The purpose of this
communication is to first identify such a Hamiltonian and then
give a complete solution of the problem in terms of the confluent
hypergeometric function.
Such a Hamiltonian as we will demonstrate below may be interpreted to represent a momentum-dependent
effective mass quantum system as guided by the choice of an underlying variable parameter $\eta$ .
While the complementary quantum problem
of the position-dependent mass has received considerable attention due to its relevance in
describing the dynamics of electrons in problems of compositionally graded crystals \cite{GKo},
quantum dots \cite{SL} and liquid crystals \cite{BGHN}, interest in momentum-dependent mass problems is a
somewhat recent curiosity arising from the observation that the parity-time symmetric Li\'enard type nonlinear
oscillator can afford complete solvability in a momentum space description \cite{RSL}. Motivated by
such a revelation we undertake a complete treatment of the Lagrangian description of (1) employing
the Jacobi Last Multiplier (JLM) approach and then moving on to a Hamiltonian formulation.
As it will turn out, the general form of the Lagrangian is guided by two choices of $\eta$ one of which has
already been studied. The second one yields a new candidate for the Hamiltonian of (1) an inquiry
of which is one of the objectives of the present work.

\section{A Lagrangian description of the Li\'{e}nard equation}

Towards this end it is instructive to review the results already
available in the literature \cite{RSL,MRS, GCGK, NT} for a generalized
class of Li\'{e}nard equation given by
\be\label{e2}\ddot{x}+f(x)\dot{x}+g(x)=0,\ee which reduces to
(\ref{e1}) for the following specific forms of $f$ and $g$:
\be\label{e3}f(x)=kx,\;\;g(x)=\omega^2 x+\frac{k^2}{9}x^3.\ee Note
that both $f$ and $g$ are odd function of $x$. To study the
dynamical aspects of (\ref{e2}) it is convenient to adopt the
method of the Jacobi Last Multiplier (JLM) whose relationship with
the Lagrangian, $L=L(t,x,\dot{x})$, for any second-order equation,
$\ddot{x}=F(t,x,\dot{x})$, is  given by \cite{ NL2, GCGK}\be
\label{k4} M=\frac{\partial^2 L}{\partial \dot{x}^2}. \ee
 $ M=M(t,x,\dot{x}) $, the JLM, satisfies
the following equation \be \label{k5} \frac{d}{dt}(\log M)+
\frac{\partial F}{\partial \dot{x}}=0.\ee In the present case
(\ref{k5}) reads \be \label{k6} \frac{d}{dt}(\log M)-f(x)=0, \ee
whose formal solution in terms of a nonlocal variable  $u$ is
\be\label{k3.8} M(t,x)=\exp\left(\int f(x) dt\right):=u^{\eta},
\ee
 where $\eta$ is a variable parameter. From (\ref{k3.8}) it is obvious that,
$\eta\dot{u} = uf(x)$ and we assume equation (\ref{e2}) can be recast
as a pair of coupled first-order
differential equations
\be\label{k1a}\dot{u}=\frac{1}{\eta}u\;f(x),\;\;\;\dot{x}=u+W(x).\ee
The functional form of $W(x)$ may be determined by differentiating the latter
equation of (\ref{k1a}) with respect to $t$ and using the former to eliminate
the nonlocal variable $u$, which yields
$$
\ddot{x} - \big(\frac{1}{\eta}f(x) + W^{\prime}(x) \big) \dot{x} + \frac{1}{\eta}Wf(x) = 0.
$$
A comparison with (\ref{e2}) shows that
$W=\eta g/f$  and the functions $f$ and $g$ are subject to
the  following constraint involving $\eta$.
  \be\label{cond1}\frac{d}{dx}\left(\frac{g}{f}\right)
  =-\frac{1}{\eta}\left(\frac{1}{\eta}+1\right)f(x).\ee
The ratio $g/f$ can be integrated out to
\be\label{g1}\eta\frac{g}{f}=\left[-\frac{\eta+1}{\eta}\int^xf(s)ds+\nu\right],\ee
 where $\nu$ is a constant of integration. It is interesting to
 note here that when  (\ref{g1}) is coupled with the choices for
 $f$ and $g$ in (\ref{e3}) it leads to a quadratic equation determining the parameter $\eta$, viz. $2\eta^2 + 9\eta + 9 = 0$,
and yields two plausible
 solutions for $\eta$:
\be\label{e4}\eta = -3\;\;\;\mbox{and}\;\;\; -3/2,\ee with $\nu =
-3 \omega^2/k$  and $-3 \omega^2/2k$ respectively. While $\eta=-3$
was investigated in \cite{RSL}, the second solution $\eta=-3/2$ is
new and is the point of focus in this paper.\\

Turning  to (\ref{k1a}) we have from (\ref{k4}) and (\ref{k3.8})
\bel{e4a}\frac{\partial^2L}{\partial
\dot{x}^2}=\left(\dot{x}-\eta\frac{g}{f}\right)^\eta,\ee which
works out to the following explicit form for $L$:
\be\label{Lag1a}L(x,\dot{x},t)=\frac{\left(\dot{x}-\eta\frac{g}{f}\right)^{\eta+2}}{(\eta+1)(\eta+2)}
+h_1(x,t)\dot{x}+h_2(x,t).\ee  In (\ref{Lag1a})  $h_1(x,t)$ and
$h_2(x,t)$ are arbitrary functions. However, consistency with
(\ref{e2}) demands that $h_1$ and $h_2$ satisfy the constraint
$h_{1t}-h_{2x}=0$. Hence there exists an auxiliary function
$G(x,t)$ in terms of which $h_1$ and $h_2$ are expressed by
 $h_1(x,t)=G_x$ and $h_2(x,t)=G_t$. As a result the
Lagrangian (\ref{Lag1a}) assumes the form \be\label{Lag1c}
L=\frac{\left(\dot{x}-\eta\frac{g}{f}\right)^{\eta+2}}{(\eta+1)(\eta+2)}
+\frac{dG}{dt},\ee where the total derivative term can be
discarded without loss of any generality. \\

\section{Hamiltonian formulations of the Li\'enard equation}

The conjugate momentum corresponding to $L$ is then defined
through \be\label{ex1} p=\frac{\partial L}{\partial
\dot{x}}=\frac{\left(\dot{x}-\eta\frac{g}{f}\right)^{\eta+1}}{(\eta+1)},\ee
implying $\dot{x}=\eta g/f+\left((\eta+1)p\right)^{1/(\eta+1)}$.
The associated  Hamiltonian, $H$,  using the standard Legendre
transformation, turns out to be \be\label{Hamlie}
H=p\dot{x}-L=\eta p\frac{g}{f}
+\frac{(\eta+1)^{\frac{\eta+2}{\eta+1}}}{(\eta+2)}
p^{\frac{\eta+2}{\eta+1}}.\ee
We can also express $H$ in terms of  a  scaled variable,
$\tilde{p}=(\eta+1)p$, whence (\ref{Hamlie})
 reads
 \be\label{Hamlie2}H(x,\tilde{p}, \eta)=\frac{1}{\eta+2}\tilde{p}^{\frac{\eta+2}{\eta+1}}
 +\frac{\eta}{\eta+1}\tilde{p}\frac{g}{f}.\ee
Equation (\ref{Hamlie2}) stands as the Hamiltonian for the
generalized Li\'{e}nard equation (\ref{e2}). Numerous models
follow from it depending on the specific choices of the functional
ratio $g/f$. The latter in turn acquires its form from the
knowledge of $f$ by solving (\ref{g1}). However, we will be
interested here in a quadratic representation of $g/f$ namely $g/f
= ax^2+b$, $a$ and $b$ are constants, to make a connection to
\cite{RSL} transparent. This is also clear from the choices of $f$
and $g$ provided in (\ref{e3}). We thus see that the two solutions
of $\eta$ furnished in (\ref{e4}) produce the following
Hamiltonians for the Li\'{e}nard oscillator (\ref{e1}):
 \bel{e5}H(x, \tilde{p},\eta=-3)= \frac{x^2}{2(3a\tilde{p})^{-1}}
 +\frac{3}{2}b\left(\sqrt{\tilde{p}}-\frac{1}{3b}\right)^2-\frac{1}{6b}.\ee
\bel{e6}H(x,\tilde{p},\eta=-3/2)=(3a\tilde{p})x^2+3b\tilde{p}+\frac{2}{\tilde{p}}.\ee
In both the above equations the run of $\tilde{p}$ is restricted
to $0 < \tilde{p} < \infty$. We observe that (\ref{e5})
corresponds to the non-standard scenario studied in \cite{RSL}
that reflects the harmonic oscillator problem.
On the other hand, (\ref{e6}) is a candidate for another
legitimate Hamiltonian associated with the Li\'{e}nard oscillator
(\ref{e1}) which also shares with (\ref{e5}) a similar feature of
interchange of the roles of variables $x$ and $\tilde{p}$. Indeed
it has the form \be\label{ex10}H=\frac{x^2}{2m(\tilde{p})}+
U(\tilde{p})\ee where from (\ref{e6}) we readily identify the mass
and potential function to be
 \bel{e8}
m(\tilde{p})=(6a\tilde{p})^{-1}\;\;\mbox{and}\;\;\;U(\tilde{p})=3b\tilde{p}+\frac{2}{\tilde{p}}.\ee
Such a system supports the following periodic solution of
$\tilde{p}$ corresponding to $x(t)$ in (\ref{ex3})
\be\label{ex2}\tilde{p}=\frac{[1-\frac{kA}{3\omega}\cos(\omega
t+\delta)]}{\sqrt{\frac{3\omega^2}{2k}-\frac{kA^2}{6}}},\ee with
$-1<kA/3\omega<1$.  The trajectory is confined to the upper-half
of the $(x-\tilde{p})$ and has the form
\be\label{ex4}\left[1+\left(\frac{k
x}{3\omega}\right)^2\right]\bar{p}^2-2\bar{p}+\left[1-\left(\frac{k
A}{3\omega}\right)^2\right]=0,\;\;\;|A|<\frac{3\omega}{k},\ee
where we have set
$\bar{p}:=\tilde{p}\sqrt{\frac{3\omega^2}{2k}-\frac{kA^2}{6}}$ and
$ |x|\leq {A}/{\sqrt{1-\left(\frac{kA}{3\omega}\right)^2}}$.

In Sec.IV, we consider the Schrodinger equation in the
presence of the momentum-dependent mass function and potential
given respectively by (\ref{e8}). We shall see that we run into an
isotonic potential \cite{RSL} in a momentum-dependent mass
background.

\section{The Schr\"{o}dinger equation with a momentum dependent
mass}

In this section we will be specifically concerned with the
quantized version of (\ref{ex10}) and seek a solution of the
corresponding Schr\"{o}dinger equation having momentum-dependent
mass and the potential function given in (\ref{e8}) in contrast to
the coordinate-dependent mass situation that has been well studied
in the literature \cite{BB, CRS123,  CCNN23, Must, GCG} in the
configuration space. In fact taking cue from such investigations
we begin this section with a von Roos type of decomposition
\cite{vR} for the generic Hamiltonian in the momentum space
 \bel{x.6}H(\hat{x}, \hat{\tilde{p}})=
 \frac{1}{4}\left[m^\alpha(\hat{\tilde{p}})\hat{x}m^\beta(\hat{\tilde{p}})\hat{x}m^\gamma(\hat{\tilde{p}})+
m^\gamma(\hat{\tilde{p}})\hat{x}m^\beta(\hat{\tilde{p}})\hat{x}m^\alpha(\hat{\tilde{p}})\right]+U(\hat{\tilde{p}}).\ee
Here $\alpha, \beta$ and $\gamma$ are the so called ambiguity
parameters which must satisfy the constraint
$\alpha+\beta+\gamma=-1$ to ensure dimensional consistency.
Following the standard procedure we apply the following
quantization rule on $\hat{x}$ and $\hat{\tilde{p}}$ in the
momentum space, noting that their corresponding operator
representations reverse their roles apart from a change in sign
compared to what we normally encounter in the configuration space
of standard quantum mechanics: \be\label{ex11}\hat{x}\rightarrow
i(\eta +1)\frac{\partial}{\partial\tilde{p}},
\;\;\hat{\tilde{p}}\rightarrow \tilde{p}.\ee  Inserting this
representation into the Schr\"{o}dinger equation,
$H\psi(\tilde{p})=E\psi(\tilde{p})$,  leads us to the differential
equation
$$-\frac{(\eta+1)^2}{2m(\tilde{p})}\left[\psi^{\prime\prime}(\tilde{p})-
\frac{m^\prime(\tilde{p})}{m(\tilde{p})}\psi^\prime(\tilde{p})+
\frac{\beta+1}{2}\left(2\frac{m^{\prime 2}(\tilde{p})
}{m^2(\tilde{p})}-\frac{m^{\prime\prime}(\tilde{p})}{m(\tilde{p})}\right)\psi(\tilde{p})
+\alpha(\alpha+\beta+1)\frac{m^{\prime 2}(\tilde{p})
}{m^2(\tilde{p})}\psi(\tilde{p})\right]$$ \bel{x.7}\hskip 50 pt
+U(\tilde{p})\psi(\tilde{p})=E\psi(\tilde{p}).\ee Further
simplification can be achieved by making the following scaling
transformation, $y=\sqrt{4\tilde{p}/3a}$, using the mass function
in (\ref{e8}) which causes (\ref{x.7}) to
 reduce to
 \bel{x.9}-(\eta+1)^2\left[\frac{d^2\psi}{dy^2}+\frac{1}{y}\frac{d\psi}{dy}+
 \frac{4\alpha(\alpha+\beta+1)}{y^2}\psi\right]=(E-U(y))\psi,\ee
and implies \bel{x.10}
\frac{d^2\psi}{dy^2}+\frac{1}{y}\frac{d\psi}{dy}+
 \frac{4\alpha(\alpha+\beta+1)}{y^2}\psi+\tilde{E}\psi-\tilde{U}\psi=0,\ee
 where $\widetilde{E}=E/(\eta+1)^2$ and $\widetilde{U}=U/(\eta+1)^2$.
Note that an explicit reference to the mass function is absent in
(\ref{x.10}) as it has been scaled out. Employing a similarity
transformation \cite{DFK} \bel{x.15}
\psi(y)=\frac{\phi(y)}{\sqrt{y}},\ee  to get rid of the linear
derivative term, (\ref{x.10}) becomes (in the momentum space)
\bel{x.14a}\frac{d^2\phi}{dy^2}+
 \left[\frac{4\alpha(\alpha+\beta+1)+\frac{1}{4}}{y^2}+\tilde{E}-\tilde{U}\right]\phi=0,\;\;\;0<y<\infty.\ee
Setting $\epsilon=-4/(\alpha(\alpha+\beta+1))$ we have therefore
\bel{x.16}\frac{d^2\phi}{dy^2}+\left(\widetilde{E}-\frac{\epsilon-\frac{1}{4}}{y^2}-\widetilde{U}\right)\phi=0,\ee
which has the standard structure of the Schr\"{o}dinger equation
for a particle of unit mass confined to an effective potential (in
momentum space)
\bel{x.17}\widetilde{U}_{eff}(y)=\widetilde{U}(y)+\frac{\epsilon-\frac{1}{4}}{y^2},\;\;\;0<y<\infty.\ee

\noindent From (\ref{e3}) we easily deduce the following values of
the constants $a$ and $b$: $a=k/9$ and $b=\omega^2/k$. When these
are inserted  into   (\ref{e8})  we have for $\eta=-3/2$ bearing
in mind the transformation $\tilde{p}=3ay^2/4$ and the fact that
   $\widetilde{U}=U/(\eta+1)^2$
 \be\widetilde{U}(y)=\omega^2
 y^2+\frac{96}{ky^2},\;\;\;0<y<\infty.\ee
Consequently (\ref{x.16}) becomes
 \bel{x.18}-\frac{d^2\phi}{dy^2}+\left[\frac{\ell(\ell+1)}{y^2}+\omega^2y^2\right]\phi=\widetilde{E}\phi,\;\;0<y<\infty\ee
where we have set $\ell(\ell+1)=\epsilon-\frac{1}{4}+96k^{-1}
>-1/4$ with $\ell$ being a real number.  The  term  in square brackets  clearly indicates the
isotonic nature of the potential. Introducing the change of
variable $y=\rho/\sqrt{\omega}$ and defining
$\Lambda=\widetilde{E}/\omega$ we find that (\ref{x.18}) may be
written as
\be\left[\frac{d^2}{d\rho^2}-\frac{\ell(\ell+1)}{\rho^2}+\Lambda-\rho^2\right]\phi(\rho)=0\ee
which under the following change of the dependent variable,
$\phi(\rho)=e^{-\rho^2/2}v(\rho)$, is  transformed to the equation
\be\left[\frac{d^2}{d\rho^2}-2\rho\frac{d}{d\rho}-\frac{\ell(\ell+1)}{\rho^2}+\Lambda-1\right]v(\rho)=0.\ee
A further transformation given by,
$v(\rho)=\rho^{\ell+1}\chi(\rho)$,  together with a change in the
independent variable, $\zeta=\rho^2$, allows us to cast the
equation into the  confluent hypergeometric form, \textit{viz}
\bel{x.19}
\zeta\frac{d^2\chi}{d\zeta^2}+\left[(\ell+\frac{3}{2})-\zeta\right]\frac{d\chi}{d\zeta}
-\left[\frac{1}{2}(\ell+\frac{3}{2})-\frac{\Lambda}{4}\right]\chi=0,\ee
which has well behaved solutions in the neighbourhood of $\zeta=0$
given by \bel{x.20}\chi(\zeta)=const.\;
_1F_1\left(\frac{1}{2}(\ell+\frac{3}{2})-\frac{\Lambda}{4};
\ell+\frac{3}{2}; \zeta \right).\ee Polynomial solutions of the
above series  results upon imposing the condition
\be\frac{1}{2}(\ell+\frac{3}{2})-\frac{\Lambda}{4}=-n,\;\;\;n=0,1,...,\ee
and leads to the equispaced energy eigenvalues
\bel{x.21}\widetilde{E}_n=\left[n+\frac{1}{2}(\ell+\frac{3}{2})\right]\omega,\;\;\;\;\;n=0,1,...\ee
for a fixed $\ell=-\frac{1}{2}\pm\sqrt{96k^{-1}+\epsilon}$. The
solution of (\ref{x.14a}) then is expressible in the form
\bel{x.22}\phi_n(y)=N_ne^{-\frac{1}{2}y^2}y^{\ell+1}\;\;_1F_1(-n;
\ell+\frac{3}{2}; y),\;\;\;0<y<\infty,\ee where $N_n$ represents
the normalization constant.\\

\section{Summary}
In this article we have considered the nonlinear Li\'{e}nard
oscillator and, adopting the JLM approach, solved completely for
the governing dynamical system. We have found that depending upon
the choice of the coupling parameters the Li\'{e}nard system has a
bi-Hamiltonian character and that for both the forms the roles of
the coordinate variable and momentum are transposed. This causes
the mass function and the potential to be explicitly
momentum-dependent. Furthermore, while one Hamiltonian is harmonic
oscillator like, the other one speaks of an isotonic potential.
While the former has been recently solved in terms of Hermite
polynomials, here we give the complete solution of the latter in
terms of a confluent hypergeometric function.

\section*{Acknowledgments}
We would like to express our sincere appreciation to Andreas Fring
and Pepin Cari\~nena for their valuable comments. The authors wish to thank
the referee for his comments which have led to an improvement of the manuscript.
This work was done while PG was visiting IH\'ES.
He would like to express his gratitude to the members of IHES for their warm hospitality.

\end{document}